\documentclass[letter,twocolumn]{IEEEtranTCOM}
\usepackage{longtable}
\usepackage{amsmath}
\usepackage{cancel}
\usepackage{amssymb}
\usepackage{graphicx}
\usepackage{longtable}
\usepackage{multirow}
\usepackage{array}
\usepackage{balance}
\usepackage[labelsep=period]{caption}
\usepackage{setspace}
\usepackage{float}
\usepackage{subfig}
\usepackage{lettrine}
\usepackage[noadjust]{cite}
\usepackage{cleveref}
\usepackage[english]{babel}
\usepackage{tikz}
\usepackage[letterpaper, left=0.625in, right=0.625in, bottom=1in, top=0.75in]{geometry}
\usepackage[figurename=Fig.]{caption}
\renewcommand{\figurename}{Fig.}

\newtheorem{proposition}{Proposition}

\newenvironment{proof}{{\textit{Proof:}}}{\hfill$\blacksquare$}

\usepackage{mathtools,amssymb,lipsum}

\begin{document}
\title{\huge A Non-Ideal NOMA-based mmWave D2D Networks with Hardware and CSI Imperfections}
\author{Leila Tlebaldiyeva, 
		Galymzhan Nauryzbayev, \IEEEmembership{Member,~IEEE,} 
		Sultangali Arzykulov, \IEEEmembership{Member,~IEEE,} 
		Yerassyl Akhmetkaziyev, 
		Mohammad S. Hashmi, \IEEEmembership{Senior~Member,~IEEE,} 
		and Ahmed M. Eltawil, \IEEEmembership{Senior~Member,~IEEE} 
		%\thanks{L. Tlebaldiyeva, G. Nauryzbayev, Y. Akhmetkaziyev, and M. S. Hashmi are with the Department of Electrical and Computer Engineering, School of Engineering and Digital Sciences, Nazarbayev University, Nur-Sultan, 010000, Kazakhstan (e-mail: \{leila.tlebaldiyeva, galymzhan.nauryzbayev, yerassyl.akhmetkaziyev, mohammad.hashmi \}@nu.edu.kz).}
		%\thanks{S. Arzykulov and A. M. Eltawil are with the Computer, Electrical and Mathematical Science and Engineering Division, King Abdullah University of Science and Technology, Thuwal, Saudi Arabia (e-mail: sultangali.arzykulov@gmail.com, ahmed.eltawil@kaust.edu.sa).}
}
\maketitle
\thispagestyle{empty}
\begin{abstract}
\boldmath
This letter investigates a non-orthogonal multiple access (NOMA) assisted millimeter-wave device-to-device (D2D) network practically limited by multiple interference noises, transceiver hardware impairments, imperfect successive interference cancellation, and channel state information mismatch. 
Generalized outage probability expressions for NOMA-D2D users are deduced and achieved results, validated by Monte Carlo simulations, are compared with the orthogonal multiple access to show the superior performance of the proposed network model.
\end{abstract}

\IEEEpeerreviewmaketitle
\section{Introduction}
\IEEEPARstart{D}{evice}-to-device (D2D) networks have received much attention as a perspective technology for the next-generation communication systems \cite{Jameel}, which enables increased coverage, resource re-use and high-rate low-latency data transmission that support many applications such as mobile cloud computing, resource sharing, {\it etc.} \cite{Ansari}. On the other hand, massive D2D communications can further escalate the spectrum congestion problem in modern radio frequency networks. At the same time, millimeter-wave (mmWave) frequency bands can be considered as a remedy able to utilize huge unlicensed spectrum bands to provide multi-Gbps transmission rates \cite{Uwa}. Thanks to its short wavelength, the mmWave band enables installation of multiple antennas on D2D devices to execute beamforming techniques for directivity and interference management purposes. For instance, the authors in \cite{Kusaladharma} modeled small-scale fading in the mmWave channel by Nakagami-$m$ distribution and used sectored antenna patterns to perform analog beamforming between D2D nodes. Another emerging technology able to address the spectrum efficiency and massive connectivity challenges for future wireless networks is a non-orthogonal multiple access (NOMA) technique, which supports users by sharing common resources such as time, code and frequency, while differentiating users based on different power levels \cite{MLiu}. One of the most attractive properties of NOMA is an easy integration with existing advanced schemes such as Internet-of-Things (IoT) \cite{MLiu}, D2D communications \cite{Wang} and mmWave systems \cite{LZhu}, {\it etc.}  For example, the authors in \cite{Wang} studied a full-duplex D2D-aided mmWave NOMA network, and closed-form expressions for the outage probability (OP) and ergodic capacity were derived. 

In contrast to these studies, this letter studies the D2D mmWave IoT-NOMA framework that simultaneously takes into account the aggregate transceiver distortions, channel and successive interference cancellation (SIC) imperfections as well as multiple unintended interference noises. The main contributions of this paper are as follows. First, a closed-form expression for the OP of the proposed NOMA-IoT network is derived and verified through Monte Carlo simulations. Second, the NOMA users' OP performance is compared against traditional orthogonal multiple access (OMA), and the advantage of NOMA is validated. Finally, the achieved results are verified by Monte Carlo simulations. 

\begin{figure}[!t]
	\centering
	\includegraphics[width=1\columnwidth]{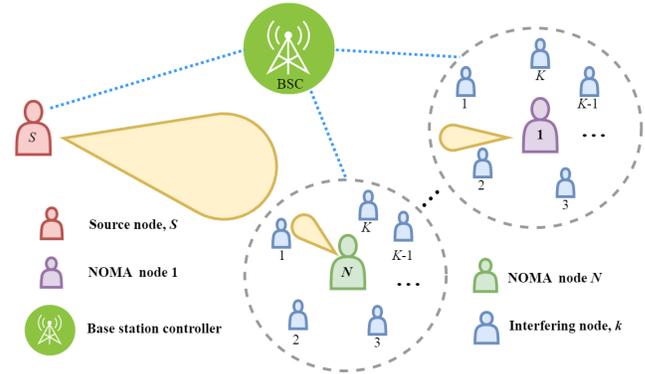}
	\caption{\small $N$-user NOMA-based mmWave IoT network.
	}
	\label{fig:draw2}
\end{figure}
	
\section{System Model}
\label{sec:system model}
Consider a downlink mmWave-based IoT network consisting of a source D2D node, denoted by $S$, and $N$ end-users with practical non-ideal transceiver hardware. $S$ aims to communicate with multiple D2D receivers, denoted by $U_n,~n\in\{1,2,\ldots,N\}$ and separated from $S$ by distances $d_n$, using a NOMA approach. Meanwhile, due to the massive users deployment, it is reasonable to assume that the users of interest are surrounded by other devices which can be clustered as depicted in Fig. \ref{fig:draw2}. For the sake of simplicity, we assume that these users create interference to the side/back lobes of the user of interest, $U_n$, through their side/back lobes; however, we consider $K$ interfering nodes per cluster for analysis purposes\footnote{Note that cooperation among the D2D and interfering nodes is maintained by a base station controller (BSC), {\it i.e.}, the D2D users are aligned along their main lobes while avoiding interference leakage from the neighboring users in the cluster.}. Moreover, small-scale fading channels in the mmWave network are modeled using a Nakagami-$m$ distribution along with analog beamforming to adequately evaluate both line-of-sight (LOS) and non-LOS (NLOS) components, which is a viable and reasonable assumption in mmWave communications \cite{leila,Heath}. Therefore, independent interfering users have fixed LOS, denoted by $m^{}_L$, or NLOS, denoted by $m^{}_N$, fading channel parameters. The corresponding path-loss exponents are equal to $\tau^{}_L$ for LOS and $\tau^{}_N$ for NLOS scenarios. Moreover, we assume all nodes to be deployed with directional antennas with steering capabilities and exploit sectored antenna pattern modeling \cite{Kusaladharma} to approximate any array pattern as
\begin{align} 
G\left(\theta\right) = \begin{cases} G_{\rm m}, &\quad |\theta |\leq \theta_{b},\\ G_{\rm s},&\quad \text {otherwise}, \end{cases}
\end{align}
where $G_{\rm m}$, $G_{\rm s}$, $\theta_{b}$ and $\theta$ denote the main and side/back lobe gains, antenna beamwidth and angle of a boresight direction, respectively. This implies that the antenna pattern has constant gain values within given main and side/back lobe sectors. For practicality reasons, the communication links are modeled assuming linear minimum mean square error as \cite{CSI}
\begin{align}
\label{csi}
h_{i} =  \tilde{h}_{i} + \epsilon, 
\end{align}
where $h_{i}$, $\tilde{h}_{i}$, and $\epsilon$ indicate the channel coefficient, its estimate and estimation error, with $\mathcal{CN}\left(0, \sigma_{\epsilon}^2\right)$, respectively, where $\sigma_{\epsilon}^2$ measures the quality of channel estimation.

According to the NOMA protocol, $S$ sends a superpositioned signal $x = \sum_{n=1}^{N}\sqrt{\alpha_n P}x_n$ to intended $N$ end-users, where $x_n$, $\alpha_n$ and $P$ stand for the message devoted for the $n$th user, the power allocation (PA) coefficient (with $\alpha_{1} > \alpha_{2} > \ldots > \alpha_{n} > \ldots > \alpha_{N}$ such that $\sum_{n=1}^{N} \alpha_{n} = 1$) and the average transmit power at $S$, respectively.

Hence, the received signal at the user of interest, {\it i.e.}, $U_i$, can be written by 
\begin{align}
\label{y}
r_i &= \sqrt{G_{\rm m}^2 d^{-\tau}_{i}}
h_i
\left(\sum\limits_{n = 1}^{N} \sqrt{\alpha_{n} P} {x_{n} + \mu_{i}} \right) 
+ \sum_{k=1}^{K}\sqrt{ G^2_{\rm s_{\it k}} d_{ik}^{-\tau} }
\nonumber \\ 
&~~~\times g_{ik} \left(\sqrt{I_k} s_{k} + \bar{\mu}_{k}\right) + w_i,
\end{align}
where $h_i$ and $g_{ik}$ denote the channel coefficients between $S$ and $U_i$ and between $U_i$ and $k$th interferer in the cluster; both coefficients are generated using independent and non-identically distributed (i.n.i.d.) random variables following Nakagami-$m$ distribution. $w_i \sim {\mathcal{CN}} \left(0,\sigma_{i}^2\right)$ denotes the additive white Gaussian noise (AWGN) term. 
The effect of residual hardware impairments (HIs) is represented by the aggregate distortion noises \cite{leila}, {\it i.e.}, $\mu_{i} \sim \mathcal{CN}\left(0,\kappa_{i}^{2}{P} \right)$ and $\bar{\mu}_{k} \sim {\mathcal {CN}}\left(0, \bar{\kappa}_{k}^{2}{I_{k}} \right)$, where $\kappa_{i}$ and $\bar{\kappa}_{k}$ indicate the compound HI levels observed in the communication links of corresponding transmitter-receiver pairs. $I_k$ stands for an average transmit power at the $k$th interferer, with $I_k = \mathbb{E}\{|s_k|^2\}$, where $s_{k}$ and $\mathbb{E}\{\cdot\}$ are the interference signal and the expectation operator, respectively.

Considering Eq. \eqref{csi} and imperfect SIC, the corresponding  signal-to-interference-noise-distortion ratio (SINDR) at $U_i$ to decode message $x_j$, $j \leqslant i$, can be written as
\begin{align}
\label{gamma}
\gamma_{j \to i} =
\frac{\alpha_{j} \rho_{i} |\tilde{h}_{i}|^{2}}{
	\rho_{i} \mathcal{A}{|\tilde{h}_{i}}|^{2} 
	+ \Sigma^{[i]} 
	+ \sum\limits_{k=1}^{K} |g_{ik}|^{2} \left(1 + \bar{\kappa}_{k}^{2}\right) \bar{\rho}_{ik}},
\end{align}
where $\rho_{i} = P G^{2}_{\rm m} d_{i}^{-\tau}$, $\bar{\rho}_{ik} = I_{k} G^{2}_{\rm s_{\it k}} d_{ik}^{-\tau}$, and $\Sigma^{[i]} = \sigma_{i}^{2}
+ \rho_{i} \left(1 + \kappa_{i}^{2}\right) \sigma_{\epsilon}^2$. $\mathcal{A} = \left(\Psi_{j} +  \tilde{\Psi}_{j} + \kappa_{i}^{2} \right)$, where $\Psi_{j} = \sum\nolimits_{t = j + 1}^{N} \alpha_{t}$, and $\tilde{\Psi}_{j} = \sum\nolimits_{l = 1}^{j - 1} {\xi_{l}}{\alpha_{l}}$, with $0 \le \xi \le 1$, describes the presence of non-ideal SIC while $\xi_{l} = 0$ and $\xi_{l} = 1$ representing perfect and imperfect SIC scenarios, respectively. Now, by defining $X^{[i]} = |\tilde{h}_{i}|^{2}$, $\zeta_{k}^{[i]} = \left(1 + \bar{\kappa}_{k}^{2}\right) \bar{\rho}_{ik}$, $Y^{[i]} = \sum_{k=1}^{K} Y^{[i]}_k = \sum_{k=1}^{K} |g_{ik}|^{2} \zeta_{k}^{[i]} $, $a_{j}^{[i]} = \alpha_{j} \rho_{i}$ and $b_{j}^{[i]} = \rho_{i} \left(\Psi_{j} +  \tilde{\Psi}_{j} + \kappa_{i}^{2} \right)$, Eq. \eqref{gamma} can be re-written as 
\begin{align}
\label{gamma1}
\gamma_{j \to i} = \frac{a_{j}^{[i]} X^{[i]}}{b_{j}^{[i]} X^{[i]} + Y^{[i]} + \Sigma^{[i]}}.
\end{align}
Note that $U_1$ decodes its own message by treating the other signals as noise and setting $\Psi_{1} = \sum_{t=2}^{N} \alpha_{t}$ and $\tilde{\Psi}_{1} = 0$, while the message $x_N$ is decoded only at $U_N$, subject to $\Psi_{N} = 0$ and perfect/imperfect SIC realizations given by $\tilde{\Psi}_{N} = \sum_{l}^{N-1} \xi_{l} \alpha_{l}$.

\section{Outage Probability}
\label{sec:Outage Analysis}
We dedicate this section to derive an exact closed-form solution for the OP under practical conditions such as hardware distortion, interference noises, and imperfect SIC/CSI scenarios. By definition, an outage is measured as the probability of a given SINDR value that falls below a certain predefined signal-to-noise (SNR) associated rate threshold, defined as $v = 2^{\mathcal{R}} - 1$, where $\mathcal{R}$ is the data rate threshold. Hence, using Eq. \eqref{gamma1}, the OP of decoding message $x_j$ by $U_i$ can be calculated as 
\begin{align}
P^{[i]}_{{\rm out},j}\left(v\right) = {\rm Pr}\left[\gamma_{j \to i} < v\right],~ 0 < j \leqslant i, i\in N.
\end{align} 
Note that the outage performance of $U_i$ is defined by $P^{[i]}_{{\rm out},i}$, as it needs to decode all messages up to $x_{j\left| j = i - 1\right.}$ to subtract their impacts before the user of interest decodes its own message. 

\begin{proposition}
Let $X^{[i]}$ be an independent non-negative Gamma distributed RV with $m_{0}$ shape and $\beta_{0}$ scale parameters, and $a_j^{[i]}$, $b_j^{[i]}$, and $\Sigma^{[i]}$ are strictly positive constants. The set of $K$ i.n.i.d. Gamma RVs, denoted by $\{Y^{[i]}_1, Y^{[i]}_2, \ldots, Y^{[i]}_K\}$, with shape ({\it i.e.}, $m_k$) and scale ({\it i.e.}, $\beta_{k}$) parameters, are generated using the probability density function (PDF) given by $f_{Y^{[i]}_k}(y) = \frac{y^{m_k - 1} e^{-\frac{y}{\beta_k}}}{\Gamma(m_k)\beta_k^{m_k}}$. Considering $K$ interfering nodes, imperfect hardware and SIC/CSI, we express the OP for the D2D-assisted mmWave NOMA network as  
{\allowdisplaybreaks
\begin{align}
\label{Pout}
P^{[i]}_{{\rm out},j}(v) &=
1 -
e^{-\frac{v \Sigma^{[i]} }{ \beta_{0} \left(a_{j}^{[i]} - b_{j}^{[i]} v\right)}}
\sum_{q=0}^{m_{0}-1} \frac{1}{\beta_{0}^{q}}
\left(\frac{v}{a_{j}^{[i]} - b_{j}^{[i]} v}\right)^{q} \nonumber\\
&~~~\times
\sum_{t=0}^{q}
\frac{\left( \Sigma^{[i]} \right)^{q-t} }{\left(q - t\right)!}
\sum_{\ell_{k} \geqslant 0, \sum_{k=1}^{K}\ell_{k}=t} 
\prod_{k=1}^{K} 
\frac{\Gamma\left(\ell_{k} + m_{k}\right)}{\ell_{k}!\Gamma\left(m_{k}\right)}
\nonumber \\
&~~\times
	\beta_{k}^{-m_{k}}
	\left(\frac{1}{\beta_{k}} + \frac{v}{ \beta_{0} \left( a_{j}^{[i]} - b_{j}^{[i]} v \right) }\right)^{- \ell_{k} - m_{k}},
\end{align}}where $a_j^{[i]} - b_j^{[i]} v > 0$, otherwise $P^{[i]}_{{\rm out},j}(v) = 1$.
\begin{proof}
Full derivation can be found in Appendix A.
\end{proof}
\end{proposition}

\section{Numerical Results}\label{sec:numerical results}
In this section, we validate our derived analytical findings by a means of Monte Carlo simulations. We define the simulation parameters as follows: the main and side/back lobe gains are set as $G_{\rm m} = 12$ dB and $G_{\rm s} = -1.1092$ dB; the average transmit power at interferers equals $I_k = 15$ dB; $v = 3$ dB; $\kappa = \kappa_{i} = \bar{\kappa}_{k}$. As mentioned earlier, due to the dense deployments of IoT devices, we assume that the users of interest are surrounded by interfering nodes that can be circled while ensuring the location of a reference receiver at the origin. For the sake of simplicity, it is also assumed that the number of interferers per cluster\footnote{Note that the interfering nodes are located at fixed locations (determined by a distance from the origin and an angle from a reference direction) in a circle with radius of $R$ and uniformly spaced concentric rings/orbits around the origin. All interferers are evenly distributed among $C$ rings, with $C = \lceil K/M \rceil$, where $M$ is the number of interferes per orbit, while assigning them with different polar angles to avoid interference blockage. For simulation purposes, we set $R = 30$ m and $M = 8$.} is fixed and equal to $K$ and two end-users are involved in the NOMA communication with a reference transmitter $S$, with corresponding distances denoted by $d_1$ and $d_2$ such that $d_1 = 2 d_2 = 100$ m. Therefore, we define the PA factors as $\alpha_{1} = 0.8$ and $\alpha_{2} = 0.2$ while generating LOS links by setting $m = m_{0} = m_k = 4$ and $\tau = 2$. 

\begin{figure}[!t]
	\centering
	\includegraphics[width=1\columnwidth]{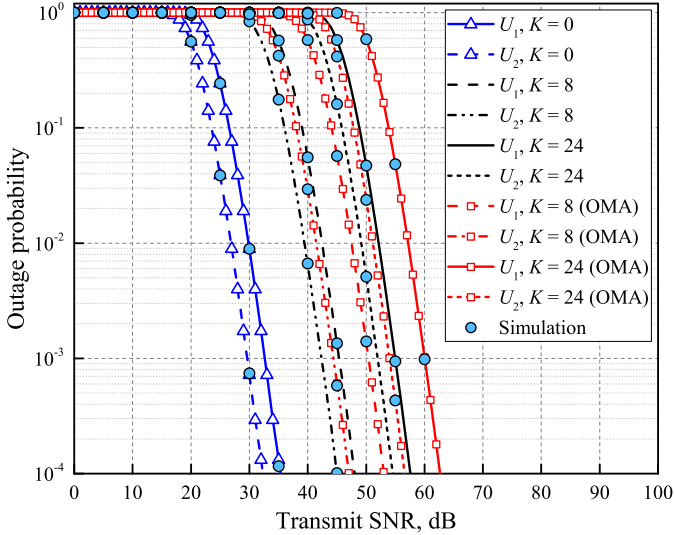}
	\caption{\small Outage probability versus transmit SNR for $K = \{0, 8, 24\}$ under ideal network conditions.}
	\label{fig:res1}
\end{figure}
In Fig. \ref{fig:res1}, we study how a number of interfering nodes, $K = \{0, 8, 24\}$, affect the OP performance of both NOMA and OMA\footnote{The OMA is regarded as a benchmark to estimate the outage performance of NOMA-aided users. For the sake of comparison fairness, the transmission rate demands of OMA are set twice as for NOMA.} users for ideal hardware and perfect SIC/CSI scenario. It is apparent that the outage performance of end-users significantly deteriorates and a corresponding gap between the curves becomes wider as the number of interferers increases. The results also reveal that the SIC-enabled end-user $U_2$ outperforms $U_1$ for given system parameters. However, subject to imperfect SIC, the performance of $U_2$ degrades when $\xi_1$ increases, as expected. Finally, one can observe that the NOMA network provides better outage than the OMA one. 

\begin{figure}[!t]
	\centering
	\includegraphics[width=1\columnwidth]{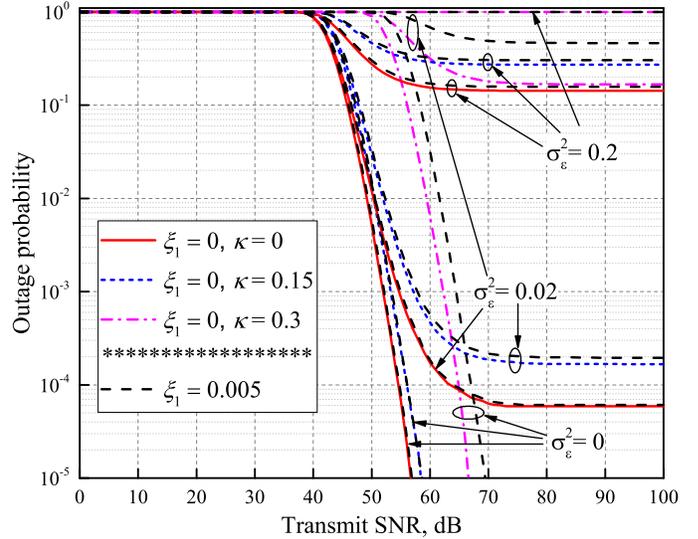}
	\caption{\small Outage probability versus transmit SNR for $K = 24$, $\kappa = \{0, 0.15, 0.3\}$, $\sigma^{2}_{\epsilon} = \{0, 0.02, 0.2\}$ and  $\xi_{1} = \{0, 0.005\}$.}
	\label{fig:res2}
\end{figure}
In Fig. \ref{fig:res2}, for the sake of figures clarity, we consider only the performance of $U_2$ with non-ideal hardware ($\kappa = \{0, 0.15, 0.3\}$) and $K = 24$ interferers under various SIC/CSI scenarios, given by $\xi_{1} = \{0, 0.005\}$ and $\sigma^{2}_{\epsilon} = \{0, 0.02, 0.2\}$. Note that the black dash lines represent the case when $\xi_{1} = 0.005$ considering all hardware/CSI realizations. It can be noticed that the non-ideality of hardware and SIC (when $\sigma^{2}_{\epsilon} = 0$) result in the outage degradation, but without its saturation that only occurs when we deal with imperfect CSI. This can be explained by the fact that both $\kappa$ and $\xi_1$ are SNR-dependent. Another useful insight is that a gap between the corresponding perfect and imperfect SIC cases becomes bigger due to the joint impact of hardware and CSI imperfections. For instance, the increase of $\sigma^{2}_{\epsilon}$ itself (consider red solid and blue short-dashed lines) does not contribute much to wider this space. At the same time, the case of perfect CSI also supports this observation, and the most noticeable change corresponds to $\kappa = 0.3$, while the other hardware realizations can be characterized by immaterial difference between the perfect and imperfect SIC curves. Finally, the system declares an outage when the considered imperfections obtain intolerable values, {\it i.e.}, $\kappa = 0.3$ and $\sigma_{\epsilon}^2 = 0.2$ for any $\xi_1$.

\section{Conclusion}\label{sec:Conclusion}
This letter studied the performance of the non-ideal NOMA-based mmWave D2D network considering the hardware, CSI, and SIC imperfections and i.n.i.d. interference noises. Closed-form analytical expressions for the OP of NOMA end-users were derived. Moreover, the proposed NOMA system model obtained better OP results compared to the OMA one (irrespective of the number of interferers), which is considered as a benchmark model. In addition, it was shown that the CSI mismatch along with non-ideal hardware drastically contribute to the outage degradation. Finally, Monte Carlo simulations validated the accurateness of the derived analytical expressions.

\begin{appendices}
\section{Proof of Proposition 1}
This section presents the derivation steps for evaluating the outage performance of the $i$th NOMA end-user in decoding $j$th message, with  $j \leqslant i$. 

Hence, the OP can be expressed using Eq. \eqref{gamma1} as
\begin{align}
\label{Pout11}
	P^{[i]}_{{\rm out},j}(v) 
	&= \Pr\left[ X^{[i]} \leq \frac{v \left(\Sigma^{[i]} + \sum\limits_{k=1}^{K} Y_{k}^{[i]} \right)}{ a_{j}^{[i]} - b_{j}^{[i]} v}
	\right]
	\nonumber\\
	= \int &
	\underset{\mathrm{\mathbb{R}^{\it K}}}{\cdots} 
	\underbrace{\left(
		\int_{0}^{\frac{v \left(\Sigma^{[i]} + \sum\limits_{k=1}^{K} y_{k}^{[i]} \right)}{ a_{j}^{[i]} - b_{j}^{[i]} v}}
		f_{X^{[i]}}(x)\textrm{d}x
	\right)}_{A_{1}} f_{\bold{Y}}(\bold{y}) \textrm{d} \bold{y},
\end{align}
where $A_1$ can be calculated using the PDF in {\it Proposition 1} as
\begin{align}
\label{A1}
A_{1} 
& =
\int_{0}^{\frac{v \left(\Sigma^{[i]} + \sum\limits_{k=1}^{K} y_{k}^{[i]} \right)}{ a_{j}^{[i]} - b_{j}^{[i]} v}}
\frac{x^{m_{0}-1} e^{-\frac{x}{\beta_{0}}}}{\Gamma(m_{0})\beta_{0}^{m_{0}}} \textrm{d} x = 1-
e^{-\frac{v \left(\Sigma^{[i]} + \sum\limits_{k=1}^{K} y_{k}^{[i]} \right)}{ \beta_{0} \left(a_{j}^{[i]} - b_{j}^{[i]} v\right)}} 
\nonumber \\
&~~~\times \sum_{q=0}^{m_{0}-1}\frac{1}{\beta_0^q \Gamma(q+1)}
\left(
\frac{v \left(\Sigma^{[i]} + \sum\limits_{k=1}^{K} y_{k}^{[i]} \right)}{ a_{j}^{[i]} - b_{j}^{[i]} v}
\right)^{q}.
\end{align}
Substituting Eq. \eqref{A1} into Eq. \eqref{Pout11}, the outage can be rewritten as 
\begin{align}
	\label{Pout1}
	&P^{[i]}_{{\rm out},j}(v) = 
	\int\limits_{0}^{\infty}\left(1-\sum_{l=0}^{m_{0}-1} 
	e^{-\frac{v \left(\Sigma^{[i]} + \sum\limits_{k=1}^{K} y_{k}^{[i]} \right)}{ \beta_{0} \left(a_{j}^{[i]} - b_{j}^{[i]} v\right)}}
	\frac{1}{\beta_0^q \Gamma(q+1) } \right.
	\nonumber\\
	&\left.\times \underbrace{\left(
		\frac{v \left(\Sigma^{[i]} + \sum\limits_{k=1}^{K} y_{k}^{[i]} \right)}{ a_{j}^{[i]} - b_{j}^{[i]} v}
		\right)^{q}}_{A_{2}} \right)
	\prod\limits_{k=1}^{K}\frac{\left(y_{k}^{[i]}\right)^{m_{k}-1}e^{-\frac{y_{k}^{[i]}}{\beta_{k}}}}{\Gamma(m_{k})\beta_{k}^{m_{k}}} \textrm{d}y_{k}^{[i]},
\end{align}
where the term $A_2$ is expanded using a binomial theorem as 
\begin{align}
	\label{A2}
	A_{2}
	&=
	\left(
	\frac{v }{ a_{j}^{[i]} - b_{j}^{[i]} v}
	\right)^{q} 
	\sum\limits_{t=0}^{q} \binom{q}{t}
	\left(\Sigma^{[i]}\right)^{q-t} 
	\left(\sum\limits_{k=1}^{K} y_{k}^{[i]} \right)^{t},
\end{align}
where $\left(\sum_{k=1}^{K} y_{k}^{[i]} \right)^{t}$ can be further expanded using a multinomial expansion as $$ \left(\sum_{k=1}^{K} y_{k}^{[i]}\right)^{t} = t!\sum\limits_{ \ell_{k} \geq 0, \sum_{k=1}^{K} \ell_{k}=t}
\left(\prod_{k=1}^{K}{\left(y_{k}^{[i]}\right)^{\ell_{k}}\over \ell_{k}!}\right).$$ 

Now, by substituting Eq. \eqref{A2} into Eq. \eqref{Pout1}, we get the following expression
\begin{align}
	\label{Pout2}
	&P^{[i]}_{{\rm out},j}(v) =
	\prod_{k=1}^{K}
	\int\limits_{0}^{\infty}\frac{\left(y_{k}^{[i]}\right)^{m_{k}-1}e^{- \left(\frac{y_{k}^{[i]}}{\beta_{k}}\right)}}{\Gamma(m_{k})\beta_{k}^{m_{k}}} 
	\textrm{d} y_{k}^{[i]}
	\nonumber\\
	&-
	e^{-\frac{v \Sigma^{[i]} }{ \beta_{0} \left(a_{j}^{[i]} - b_{j}^{[i]} v\right)}}
	\sum_{q=0}^{m_{0}-1} \frac{1}{\beta_{0}^{q}q!}
	\left(\frac{v}{a_{j}^{[i]} - b_{j}^{[i]} v}\right)^{q}
	\nonumber\\
	&\times
	\sum_{t=0}^{q}\binom{q}{t}\left( \Sigma^{[i]} \right)^{q-t} t!
	\sum_{\ell_{k} \geqslant 0, \sum_{k=1}^{K}\ell_{k}=t} \prod_{k=1}^{K}\frac{1}{\ell_{k}!\Gamma(m_{k})\beta_{k}^{m_{k}}} \nonumber\\
	&\times
	\int\limits_{0}^{\infty}\left( y_{k}^{[i]} \right)^{\ell_{k}+m_{k}-1}e^{-y_{k}^{[i]} \left(\frac{1}{\beta_{k}} + \frac{v}{ \beta_{0} \left( a_{j}^{[i]} - b_{j}^{[i]} v \right) }\right)} \textrm{d} y_{k}^{[i]}.
\end{align}
Using \cite[Eq. (8.310.1)]{grads}, we obtain the exact closed-form expression for the OP of the user of interest, as in \eqref{Pout}.
\end{appendices} 

\ifCLASSOPTIONcaptionsoff
\newpage
\fi
\balance

\end{document}